\documentclass[10pt,thmsa]{article}
\usepackage{amsfonts}

\begin{document}

\title{A comment on the Outgoing Radiation Condition for the gravitational field
and the Peeling Theorem.}
\author{Juan Antonio Valiente Kroon. \\
School of Mathematical Sciences,\\
Queen Mary \& Westfield College,\\
Mile End Road, London E1 4NS,\\
United Kingdom.}
\date{\today}
\maketitle

\begin{abstract}
The connection between the Bondi-Sachs (BS) and the Newman-Penrose (NP) 
framework for the study of the asymptotics of the gravitational field is 
done. In particular the coordinate transformation relationg the BS luminosity 
parameter and the NP affine parameter is obtained. Using this coordinate 
transformation it is possible to express BS quantities in terms of NP 
quantities, and to show that if the Outgoing Radiation Condition is not 
satisfied then the spacetime will not decay in the way prescribed by the 
Peeling theorem.
\end{abstract}

\section{Introduction.}

In the pioneering work by Bondi et al.\cite{viii} and Sachs \cite{vii} on the
asymptotic behaviour of the gravitational field of an isolated body, the 
following metric was used

\begin{equation}
ds^2=\frac{Ve^{2\beta }}{\widetilde{r}}du^2+2e^{2\beta }dud\widetilde{r}-%
\widetilde{r}^2h_{ij}(dx^i-U^idu)(dx^j-U^jdu),  \label{bondisachs}
\end{equation}

\begin{eqnarray}
h_{ij} &=&\left( 
\begin{array}{cc}
e^{2\gamma }\cosh 2\delta  & \sinh 2\delta \sin \theta  \\ 
\sinh 2\delta \sin \theta  & e^{-2\gamma }\cosh 2\delta \sin ^2\theta 
\end{array}
\right) ,  \label{angdown} \\
h^{ij} &=&\left( 
\begin{array}{cc}
e^{-2\gamma }\cosh 2\delta  & -\sinh 2\delta \csc \theta  \\ 
-\sinh 2\delta \csc \theta  & e^{2\gamma }\cosh 2\delta \csc ^2\theta 
\end{array}
\right). \label{angup}
\end{eqnarray}
A nice property of this metric is that the field equations form a hierarchy
(4 hypersurface equations, 2 standard equations, 1 trivial equation,  and  3 
supplementary conditions). So if the
the initial values of the functions $\gamma$ and $\delta$ are given on a null 
hypersurface, then it is possible to solve the hypersurface equations so that the values of
the remaining functions ($\beta ,V,$ $U^i$) on that hypersurface can be 
obtained. Then, using the evolution equations we can obtain their values for previous
retarded times. The asymptotic study of Bondi et al. and
Sachs used the following \emph{Ansatz }for $\gamma $ and $\delta $:

\begin{eqnarray}
\gamma  &=&cr^{-1}+\gamma _3r^{-3}+... , \\
\delta  &=&dr^{-1}+\delta _3r^{-3}+... .
\end{eqnarray}
The important fact to realize here is the absence of the $r^{-2}$ term in
both expansions. When solving the hypersurface equations, two integrations
with respect to $r$ will be carried out; hence if the $r^{-2}$ term is present, 
then terms of the form $r^{-i}\ln r$  will arise in the functions $V$ and 
$U^i$ (see \cite{xiv} for an example). This
Ansatz was known by the misleading name of the \emph{Outgoing Radiation
Condition of the gravitational field }(ORC), in analogy
to the Sommerfeld condition for the electromagnetic field. 

Now, the ORC does not rule out
the existence of incoming gravitational radiation travelling infinitely long 
distances (i.e. radiation coming from null infinity). In order to do so, a 
condition on the news function at past null infinity should be imposed \cite{leipold}.
The presence of incoming radiation of finite duration is not a problem, as it may 
describe the phenomena of gravitational wave scattering and gravitational wave tails
\cite{tails}, \cite{torrence}, \cite{bonnor} that die off suitably in a 
neighborhood of $\mathfrak I$.

If we keep the $r^{-2}$ terms in our expansions, then we enter into the realm of
the \emph{polyhomogeneous spacetimes}; spacetimes that can be expanded
asymptotically in a combination of powers of $1/r$ and $\ln r$. Some study
on these spacetimes has been done \cite{xiv}, \cite{chrusciel2}, \cite
{javk98a}.

Another framework for the treatment of the gravitational radiation is
the Newman-Penrose formalism \cite{n-p62}. In the NP framework we find that the  field
equations are naturally adapted for the study of the characteristic initial
value problem. The equations also form a (more lengthy) hierarchy of first
order differential equations. In this
case the initial data that has to be prescribed on the initial hypersurface is
contained in the $\Psi _0$ of the Weyl tensor. The crucial assumption in the
NP framework is that the components of the Weyl tensor fall off in the way
prescribed by the Peeling Theorem:

\begin{equation}
\Psi _k=O(r^{k-5}),
\end{equation}
and in particular the data on the initial hypersurface ($\Psi_0$) should decay
as $O(r^{-5})$ \cite{penrose}, \cite{n-p62}.
The objective of this note is to find the connection between the Bondi-Sachs
quantities (the coefficients in $\gamma$ and $\delta$) and the Newman-Penrose ones
(the coefficients in $\Psi_0$); with these tools in hand it will be
shown that the if the Outgoing Radiation Condition is not satisfied then the
Peeling theorem does not hold. 

\section{The coordinates.}

\subsection{The Bondi coordinates.}

The coordinate $u$ in the line element of equation (\ref{bondisachs}) is a 
retarded time that parametrizes outgoing null hypersurfaces. The angular 
coordinates $\theta$ and $\varphi$ are constructed in such
a way that they remain constant along the generators of the null 
hypersurfaces, and $\widetilde{r}$  is a \emph{luminosity
parameter} that satisfies 
\begin{equation}
\widetilde{r}^4\sin ^2\theta =\det h_{ij}=(\det
h^{ij})^{-1}=g_{22}g_{33}-(g_{23})^2.  \label{detbondi}
\end{equation}

\subsection{The NP coordinates.}

The coordinates used in the NP treatment can be constructed in a
similar way. The key difference lies in the choice of the radial coordinate.
Newman and Penrose use as radial coordinate the  affine parameter $r$ of the 
generators of the null hypersurfaces $u=const$. There is some freedom left in the
 
choice of this coordinate. The change

\begin{equation}
r^{\prime }=ar+b.
\end{equation}
can always be performed. The scaling of the affine parameter is chosen such 
that the contravariant metric tensor has the form

\begin{equation}
g_{NP}^{ij}=\left( 
\begin{array}{cccc}
0 & 1 & 0 & 0 \\ 
1 & 2Q & C^\theta & C^\varphi \\ 
0 & C^\theta & -2 \xi^\theta \overline{\xi}^\theta & -\xi^\theta \overline{\xi}^
\varphi - \overline{\xi}^\theta \xi^\varphi \\
0 & C^\varphi & -\xi^\theta \overline{\xi}^
\varphi - \overline{\xi}^\theta \xi^\varphi & -2 \xi^\varphi 
\overline{\xi}^\varphi
\end{array}
\right) ,
\end{equation}
while the freedom in the choice of the origin is generally used to eliminate
an arbitrary function of integration that appears in the expansion of the
spin coefficient $\rho$. The real functions $Q$, $C^{\theta}$, 
$C^{\varphi}$ and the complex functions $\xi^{\theta}$ and $\xi^{\varphi}$ 
depend on all four coordinates. We notice that in the BS treatment there are 6
metric functions whilst in the NP framework there are 7. This difference can 
be traced back to the choice of the luminosity parameter in the Bondi metric which fixes 
the form of the determinant of the angular part of the metric. The choice of 
the radial coordinate as an affine parameter is necessary in the NP formalism.
If we were to use a luminosity parameter instead, then the field equations would not 
give rise to an easy to handle hierarchy. One would have to solve all the NP field equations at once!

\subsection{ The relation between Bondi's luminosity parameter and NP affine
parameter.}

As seen before, the BS coordinates and the NP coordinate differ essentially
in the construction of the radial coordinate. The connection between the two
coordinates can be found easily by equating the determinants of the
``angular part'' of the metrics. On one hand one has 

\begin{equation}
\det h_{BS}^{ij}=\frac 1{\widetilde{r}^4\sin ^2\theta },
\end{equation}
by definition. On the other hand, the determinant of the ``angular part'' of
the NP metric is given by 

\begin{equation}
\det h_{NP}^{ij}=-\left( \xi ^\theta \overline{\xi }^\varphi -\xi ^\varphi 
\overline{\xi }^\theta \right) ^2.  \label{detnp}
\end{equation}
Imposing equality of the two determinants one obtains:

\begin{equation}
\widetilde{r}^2=\frac{-i\csc \theta }{\left( \xi ^\theta \overline{\xi }%
^\varphi -\xi ^\varphi \overline{\xi }^\theta \right) },
\label{affine-luminosity}
\end{equation}
where the right hand side is a function of the NP coordinates $(u,r,\theta
,\varphi )$. The minus sign is set in order to have $\widetilde{r}^2\geq 0$.

\section{The relation between the ORC and the Peeling Theorem.}

Although the Einstein field equations are consistent with spacetimes that
fall off as 
\begin{equation}
\Psi _0=O(r^{-3}\ln ^{N_3}r)
\end{equation}
\cite{javk98a}, for our  purposes  it is sufficient to consider a spacetime
such that

\begin{equation}
\Psi _0=\Psi _0^{4,0}r^{-4}+O(r^{-5}\ln ^{N_5}).
\end{equation}
Then using the techniques of reference \cite{javk98a} we see that 
\begin{equation}
\sigma =\sigma _{2,0}r^{-2}-\Psi _0^{4,0}r^{-3}+O(r^{-4}\ln ^{N_5}r),
\end{equation}

and
\begin{equation}
\rho =r^{-1}-\sigma _{2,0}\overline{\sigma }_{2,0}r^{-3}+\frac 12\left(
\sigma _{2,0}\overline{\Psi }_0^{4,0}+\overline{\sigma }_{2,0}\Psi
_0^{4,0}\right) r^{-4}+O(r^{-5}\ln ^{N_5}r).
\end{equation}

Whence using the commutator equations one finds

\begin{equation}
\xi ^i=\xi _0^ir^{-1}-\overline{\xi }_0^i\sigma _{2,0}r^{-2}+\left( \xi
_0^i\sigma _{2,0}\overline{\sigma }_{2,0}+\frac 12\Psi _0^{4,0}\overline{\xi 
}_0^i\right) r^{-3}+O(r^{-4}\ln ^Nr),
\end{equation}
where 
\begin{equation}
\xi _0^\theta =\frac 1{\sqrt{2}},
\end{equation}

\begin{equation}
\xi _0^\varphi =\frac{-i}{\sqrt{2}}\csc \theta,
\end{equation}
if the cuts of $\mathfrak{I\ }$ are chosen to be $S^2$ metrically.

The substitution of these expansions into equation (\ref{affine-luminosity}) 
yields the transformation linking the NP affine parameter and the BS 
luminosity parameter.

\begin{equation}
\widetilde{r}=r-\frac 12\sigma _{2,0}\overline{\sigma }_{2,0}r^{-1}+\frac
16\left( \sigma _{2,0}\overline{\Psi }_0^{4,0}+\overline{\sigma }_{2,0}\Psi
_0^{4,0}\right) r^{-2}+...,
\end{equation}
so that the metric functions $\gamma$ and $\delta$ can be written in terms of 
the NP quantities as:

\begin{equation}
\gamma =\frac 12(\overline{\sigma }_{2,0}+\sigma _{2,0})\widetilde{r}%
^{-1}-\left( \overline{\Psi }_0^{4,0}+\Psi _0^{4,0}\right) \widetilde{r}%
^{-2}+...,
\end{equation}

\begin{equation}
\delta =\frac i2(\overline{\sigma }_{2,0}-\sigma _{2,0})\widetilde{r}%
^{-1}-i\left( \overline{\Psi }_0^{4,0}-\Psi _0^{4,0}\right) \widetilde{r}%
^{-2}+....
\end{equation}
We see that the term that breaks the peeling behaviour ($\Psi _0^{4,0}$)
gives rise to the coefficients forbidden by the outgoing radiation
condition. A similar study can be carried out for more general polyhomogeneous
space-times giving as a result that the coefficients $\Psi _0^{3,k}$ are
related to logarithmic terms in the $1/r$ terms of $\gamma $ and $\delta $,
etc. In principle these expansions can be performed up to any desired order.

\section{Conclusions.}

As we have seen, the Outgoing Radiation Condition and the $\Psi_0 = O(r^{-5})$
condition of Penrose are closely related. If the ORC is not satisfied then 
Penrose's condition will not be satisfied. However, the two conditions are not 
completely equivalent. Penrose's condition is stronger. For example, if we 
consider a polyhomogeneous spacetime such that 

\begin{equation}
\gamma = cr^{-1} + \gamma_{3}r^{-3} + \gamma_{31}r^{-3}\ln r +... ,
\end{equation}

\begin{equation}
\delta = dr^{-1} + \delta_{3}r^{-3} + \delta_{31}r^{-3}\ln r +... .
\end{equation}

\noindent then clearly, the ORC will be satisfied, but the leading behaviour of $\Psi_0$
will go as $O(r^{-5}\ln r)$. The null infinity for this spacetime will not be 
smooth, as the conformally rescaled $\Psi_0$ will blow up there.

\section*{Acknowledgments.}

I would like to thank my supervisor  Prof. M A H MacCallum for a lot of
discussions and encouragement. Also to Dr. M Mars for related discussions.
The author has a grant (110441/110491 ) from the Consejo Nacional de Ciencia
 y Tecnolog\'{\i}a (CONACYT), Mexico.

\end{document}